\documentclass[aps,twocolumn,floatfix,superscriptaddress,amsmath,showpacs,showkeys,prb]{revtex4-1}

\usepackage{t1enc}
\usepackage[final]{graphicx}
\usepackage{graphicx}
\usepackage{epsfig}
\usepackage{bm}
\usepackage[normalem]{ulem}
\usepackage{color}
\usepackage{float}
\usepackage{url}
\usepackage{hyperref}

\begin{document}

\title{Magnetization reversal in mixed ferrite-chromite perovskites with  non magnetic cation on the A-site}

\author{Orlando V. Billoni}
\email{billoni@famaf.unc.edu.ar}
\affiliation{Facultad de Matem\'atica, Astronom\'{\i}a, 
F\'{\i}sica y Computaci\'on, Universidad Nacional de C\'ordoba and \\ Instituto de
F\'{\i}sica Enrique Gaviola  (IFEG-CONICET), Ciudad Universitaria,
5000 C\'ordoba, Argentina.}

\author{Fernando Pomiro}
\email{fernandopomiro@gmail.com}
\affiliation{INFIQC (CONICET -- Universidad Nacional de C\'ordoba), Departamento de Fisicoqu\'{\i}mica, Facultad de Ciencias Qu\'{\i}micas, Universidad Nacional 
de C\'ordoba, Haya de la Torre esq. Medina Allende, Ciudad Universitaria, X5000HUA C\'ordoba, Argentina.}

\author{Sergio A. Cannas}
\email{cannas@famaf.unc.edu.ar}
\affiliation{Facultad de Matem\'atica, Astronom\'{\i}a, 
F\'{\i}sica y Computaci\'on, Universidad Nacional de C\'ordoba and \\ Instituto de
F\'{\i}sica Enrique Gaviola  (IFEG-CONICET), Ciudad Universitaria,
5000 C\'ordoba, Argentina.}

\author{Christine Martin}
\email{christine.martin@ensicaen.fr}
\affiliation{Laboratoire CRISMAT, UMR 6508 CNRS/ENSICAEN/UCBN, 6 Boulevard Marechal Juin, 14050 Caen cedex,
France.}

\author{Antoine Maignan}
\email{antoine.maignan@ensicaen.fr}
\affiliation{Laboratoire CRISMAT, UMR 6508 CNRS/ENSICAEN/UCBN, 6 Boulevard Marechal Juin, 14050 Caen cedex,
France.}

\author{Raul E. Carbonio}
\email{rcarbonio@gmail.com}
\affiliation{INFIQC (CONICET -- Universidad Nacional de C\'ordoba), Departamento de Fisicoqu\'{\i}mica, Facultad de Ciencias Qu\'{\i}micas, Universidad Nacional 
de C\'ordoba, Haya de la Torre esq. Medina Allende, Ciudad Universitaria, X5000HUA C\'ordoba, Argentina.}

\date{\today}

\begin{abstract}
In this work, we have performed Monte Carlo simulations in a classical model for RFe$_{1-x}$Cr$_x$O$_3$ with R=Y and Lu, 
comparing the numerical simulations with experiments and mean field calculations. In the analyzed compounds, 
the antisymmetric exchange or Dzyaloshinskii-Moriya (DM) interaction induced a weak ferromagnetism due to a 
canting of the antiferromagnetically ordered spins. This model is able to reproduce the magnetization reversal 
(MR) observed experimentally in a field cooling process for intermediate $x$ values and the dependence with $x$ of 
the critical temperatures. We also analyzed the conditions for the existence of MR in terms of the strength of 
DM interactions between Fe$^{3+}$ and Cr$^{3+}$ ions with the x values variations. 
\end{abstract}




\maketitle
\section*{Introduction} \label{intro}

Some magnetic systems when cooled in the presence of low magnetic fields show magnetization reversal (MR).
At high temperatures the magnetization  points in the direction of the applied field while at
a certain temperature the magnetization reverses, becoming opposite to the magnetic field in a low
temperature range.
In particular, this phenomenon has been observed in orthorhombic (space group: \emph{Pbnm}) perosvkites
like RMO$_{3}$ with R=rare earth or yttrium and M=iron, chromium or vanadium \cite{Kadomtseva77JETP, Yoshii00JSSC, Yoshii01aJSSC, Yoshii01bJSSC, Mao11APL, Dasari12EPL, Mandal13JSSC, Ren1998}.
These materials exhibit a weak ferromagnetic behavior below the N\'{e}el temperature (T$_{N}$),
arising from a slight canting of the antiferromagnetic backbone. The weak ferromagnetism (WFM) observed
in these compounds can be due to two mechanisms related with two different  magnetic interactions: antisymmetric exchange
or Dzyaloshinskii-Moriya interaction (DM) and single-ion magnetocrystalline anisotropy \cite{Treves62PR, Moriya1960}.
In particular, in orthochromites RCrO$_3$ and orthoferrites RFeO$_3$ the WFM is due mainly to DM interactions \cite{Treves62PR}.

MR was also observed in several ferrimagnetic systems such as spinels \citep{Gorter1953,Menyuk1960},
garnets \citep{Pauthenet1958}, among others. In these materials, MR has been explained by a different
temperature dependence of the sublattice magnetization arising from different crystallographic sites,
as predicted by N\'{e}el for spinel systems. However, this explanation cannot be applied to the
orthorhombic perovskites with formula RM$_{1-x}{\rm M}'_x$O$_{3}$ where R is a nonmagnetic ion (for example Y$^{3+}$ or Lu$^{3+}$),
because the magnetic ions occupy a single crystallographic site.
In the case of YVO$_{3}$, the origin of MR has been explained based on a competition between DM interaction
and single-ion magnetic anisotropy \citep{Ren2000}.

Some years ago, the presence of MR was also reported in polycrystalline perovskites with two magnetic
transition ions randomly positioned at the B-site and non magnetic R cation at the A site. Some examples
are BiFe$_{0.5}$Mn$_{0.5}$O$_{3}$, LaFe$_{0.5}$Cr$_{0.5}$O$_{3}$, YFe$_{0.5}$Cr$_{0.5}$O$_{3}$ and
LuFe$_{0.5}$Cr$_{0.5}$O$_{3}$ \citep{Mao2011,Mandal2010,Azad2005,Pomiro2016}.
In a work by Kadomtseva  et al. [\onlinecite{Kadomtseva77JETP}] the DM interactions were 
successfully used to explain the anomalous magnetic properties of single-crystal  YFe$_{1-x}$Cr$_x$O$_3$ with different Cr contents. They showed that these compounds
are weak ferrimagnets with a mixed character of the DM interaction. 
Moreover, the competing character of DM interactions is used in a mean field (MF) approximation by
Dasari et al. [\onlinecite{Dasari12EPL}] to explain the field cooling curves of polycrystalline
YFe$_{1-x}$Cr$_x$O$_3$ for $0\le x \le 1$. In their work the dependence
of magnetization as a function of temperature, for the entire range of composition, is explained from the
interplay of DM interactions of the Fe--O--Fe, Cr--O--Cr and Cr--O--Fe bounds. At intermediate
compositions (x=0.4 and 0.5) MR is also reported in this work.

Numerical simulations have been proved to be useful to model magnetic properties of perovskites.
Several studies of magnetic perovskites have been performed using Monte Carlo simulations (MC) \cite{Murtazaev05LTP,RestrepoParra10JMMM,RestrepoParra11JMMM},
for instance, to characterize the critical behavior  in yttrium orthoferrites \cite{Murtazaev05LTP} and
in La$_{2/3}$Ca$_{1/3}$MnO$_{3}$ \cite{RestrepoParra10JMMM,RestrepoParra11JMMM}. However,
to the best or our knowledge, MR has not been studied using MC simulations. In the case of solid solutions,
MC simulations can take into account fluctuations in the distribution of atomic species
and thermal fluctuation that cannot be considered in mean field models.

In this work we have performed MC simulations using a classical model for RFe$_{1-x}$Cr$_{x}$O$_3$
with R = Y or Lu, comparing the numerical simulations with experiments and mean field  calculations~\cite{Dasari12EPL,Hashimoto63JPSJ}.
We also adapted MF approximations to test our MC simulations. 
This model is able to reproduce the magnetization reversal (MR) observed in a field cooling process
for intermediate $x$ values and the dependence on $x$ of the critical temperature. 

We also  analyzed the conditions for the existence
of MR in terms of the strength of DM interactions between Fe$^{3+}$  and Cr$^{3+}$ and the chromium content.

\section{Methods}
\label{methods}
Neutron diffraction studies have shown that the magnetic structure of RFe$_{1-x}$Cr$_x$0$_3$ compounds
with a non-magnetic R ion (space group :$Pbnm$) is $\Gamma_4(G_x,A_y,F_z)$ in the Bertaut notation \cite{Bertaut}.
In this structure the moments are oriented mainly in an AFM type-G arrangement along the x-direction. 
A nonzero ferromagnetic component along the z-axis (canted configuration) and an AFM type-A arrangement
along the y-axis are allowed by symmetry \citep{Treves62PR,Sherwood1959}.
 
We model the RFe$_{1-x}$Cr$_x$O$_3$  perosvkites, with R$=$Lu or Y using the following Hamiltonian of 
classical Heisenberg spins  lying in the nodes of  a cubic lattice with $N = (L\times L \times L)$ sites,

\begin{eqnarray}
\mathcal{H} &=& - \frac{1}{2} \sum_{\langle i,j \rangle }[J_{ij} \vec{S}_i \cdot \vec{S}_j +  \vec{D}_{ ij}\cdot (\vec{S}_i \times \vec{S}_j)] \\ \nonumber
            & &- \sum_{i} K_{i} (S_i^x)^2 - H \sum_{i} m_{i} S_i^z, 
\end{eqnarray}

\noindent where $\langle ... \rangle$ means a sum over the nearest neighbor sites and $\vec{S}_i$ are unitary vectors. 
$J_{ij} < 0$ takes into account the superexchange interaction and $\vec{D}_{ij}$ the anti-symmetric Dzyalshinskii-Moriya 
interactions, where this vector points in the $\hat{j}$ direction. Due to the collective tilting of the (Fe,Cr)O$_6$ octahedra 
the DM interaction is staggered. $H$  correspond to the external applied field and is expressed as, $H = B \mu_{Fe}/k_B $, where
$B$ is the external field and $\mu_{Fe} = g \mu_B S_{Fe}$ with $g=2$ the gyromagnetic factor constant,  $\mu_B$ the Bohr 
magneton and $S_{Fe} = 5/2$ is the total spin of Fe$3+$ ion --equivalently $S_{Cr} = 3/2$ for  Cr$3+$ ion.  
Then,  $m_{i} = 1$ for then Fe$^{3+}$ ions and $m_{i} = S_{Cr}/S_{Fe} = 0.6$ for the Cr$^{3+}$ ions.  
Both interactions, $J_{ij}$  and $D_{ij}$, depend on the type of ions (Fe$^{3+}$ or Cr$^{3+}$) that occupy sites $i$ and $j$, so each 
pair interaction can take three different values. Suppose that site $1$ is occupied by Cr$^{3+}$ and site $2$ by Fe$^{3+}$  ion, then  
the super-exchange interaction couplings are $J_{22} = 2 S_{Fe}^2J_{FeFe}/k_B$, 
$J_{12}=J_{21} =  2 S_{Fe}S_{Cr}J_{FeCr}/k_B$ and $J_{11}=2 S_{Cr}^2J_{CrCr}/k_B$, where $J_{\alpha \beta}$ 
-- with $\alpha, \beta  =$ Cr or Fe  -- are the exchange integrals.  
In the case of DM interactions $D_{22} = S_{Fe}^2D_{FeFe}/k_B$; $-D_{12} =  D_{21}  = S_{Fe} S_{Cr} D_{FeCr}/k_B$ 
and $D_{11} = S_{Cr}^2 D_{CrCr}/k_B$. Finally, the single site interactions corresponding to the uniaxial anisotropy are 
$K_{1} = S_{Cr} K_{Cr}/k_B=K_{2} = S_{Fe} K_{Fe}/k_B >0$ which point in the $x$ direction. $k_B$ is the Boltzmann constant. 
For simplicity we consider the same anisotropy for Cr$^{3+}$ and Fe$^{3+}$ ions. 

\subsection{Monte Carlo methods}
\label{MC}

We performed Monte Carlo simulations using a Metropolis algorithm. Along this work we considered  
a cubic lattice with $N= 40 \times 40 \times 40$ sites and open boundary conditions. In order to simulate RFe$_{1-x}$Cr$_x$O$_3$ compounds   
the sites of the cubic lattice are occupied by Cr$^{3+}$ ions with probability $x$, and with probability $(1-x)$ by the Fe$^{+3}$ ions.  
Since all the super-exchange interactions are antiferromagnetic the system can be divided into two sublattices $A$ and $B$,  
each one ferromagnetically ordered in the $\hat{i}$ direction and opposite to the other. We computed the sublattice magnetization, 
\begin{equation}
\vec{m}_{\alpha} = \frac{1}{N} \sum_{\vec{S}_{i} \in \{ \vec{S}_{\alpha} \}} \vec{S}_i, 
\end{equation}
\noindent where $\{ \vec{S}_{\alpha} \}$ with $\alpha=A,B$ is the set of spins belonging to sublattice $A$ or
sublattice $B$, and the susceptibility,
\begin{equation}
\chi_{\alpha}= (\frac{N}{k_B T})(\langle m_{\alpha}^2 \rangle - \langle m_{\alpha} \rangle^2 ),
\end{equation}
\noindent where $\langle ... \rangle$ means a thermal average. At each temperature we
equilibrated the system using $10^5$ Monte Carlo steps (MCS). After that we get  the thermal averages using
another $10^5$ MCS, measuring the quantities (e.g. the magnetization) every $10^2$ MCS.  
From the peak of the susceptibility  we obtained the critical temperature as function
of the Chromium content for $x=0, 0.1, 0.2, ..., 1$. 
As we will explain later, the values of the $J_{11}$ and $J_{22}$ interactions were chosen in order
to reproduce the N\'eel temperature $T_{N}$ of the pure compounds, RCrO$_3$ and RFeO$_3$, respectively. 
 $J_{12}$ were considered as a free parameter,  to be fitted from the experiments. 
The value used for  $K_1 = K_2$ in all the simulations was  $K_1= 7\times 10^{-3} J_{22}$~\cite{Murtazaev05LTP}. 
In the case of LuFe$_{1-x}$Cr$_{x}$O$_3$ we used the following values for the  DM interactions 
$D_{11}=0.74 \times 10^{-2} J_{22}$ and $D_{22}= 2.14 \times 10^{-2} J_{22}$, taken from
Refs.[\onlinecite{Hornreich76PRB}] and [\onlinecite{Treves65JAP}], respectively. 
There is no estimation of $D_{12}$ in the literature, so we assume as a reference the value $D_{12} = -1.7\times 10^{-2}J_{22}$  
to obtain the critical temperatures, considering that a similar value was obtained by  Dasari et al\cite{Dasari12EPL} 
fitting YFe$_{1-x}$Cr$_x$O$_3$  data.
Since DM interactions are considerably lower than super-exchange interactions, small variations
of this interactions does not substantially affect the antiferromagnetic ordering temperatures.

\subsection{Effective model}
\label{EM}
\begin{figure}
\begin{center}
\includegraphics[scale=0.75, angle=00]{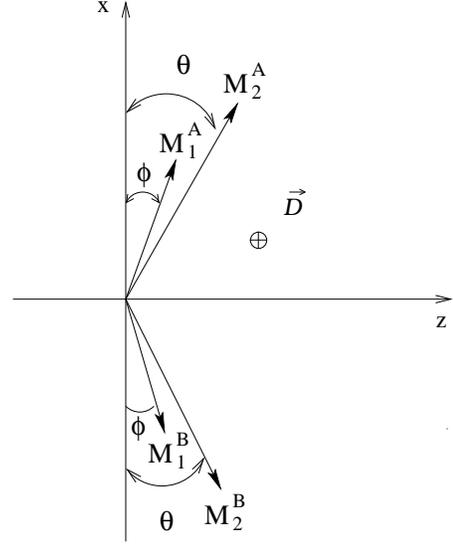}
\caption{\label{fig0}  
Sketch of the configuration of the two sublattice magnetization. 
}
\end{center}
\end{figure}

In order to get a deeper physical insight about the low temperature behavior of these systems, we compared the MC 
results against an effective  model that generalizes some ideas introduced by Dasari et al\cite{Dasari12EPL}.
In this model a site $i$ is occupied with probability $P_1[x] = P_{Cr}[x]=x$ by a Cr$^{3+}$ ion and with 
probability $P_2[x] = P_{Fe}[x] = (1-x)$ by a Fe$^{3+}$ ion.
The energy, in a two-sublattice approximation, is then given by 

\begin{eqnarray}
\label{eq-mf}
\nonumber
E & = & z[\mathcal{J}_{11} \vec{M}_1^A \cdot \vec{M}_1^B + \mathcal{J}_{22} \vec{M}_2^A \cdot \vec{M}_2^B  + \\ \nonumber 
  & + &\mathcal{J}_{12}( \vec{M}_1^A \cdot \vec{M}_2^B + \vec{M}_2^A \cdot \vec{M}_1^B) + \\ \nonumber 
  & + & \mathcal{D}_{11}\cdot( \vec{M}_1^A \times \vec{M}_1^B) +  \mathcal{D}_{22} \cdot (\vec{M}_2^A \times \vec{M}_2^B) + \\ \nonumber 
  & + & \mathcal{D}_{12} \cdot (\vec{M}_1^A \times \vec{M}_2^B + \vec{M}_2^A \times \vec{M}_1^B)] + \\ \nonumber 
  & - & \mathcal{K}_1 [(M_{1x}^A)^2 + (M_{1x}^B)^2] - \mathcal{K}_2 [(M_{2x}^A)^2 + (M_{2x}^B)^2)] + \\  
  & - & \mathcal{H}_1 [M_{1z}^A + M_{1z}^B] - \mathcal{H}_2 [M_{2z}^A + M_{2z}^B],  
\end{eqnarray}
here $\vec{M}_1^{\alpha}$, with $\alpha = A$ or $B$ is the total magnetization of the chromium ions belonging to the  
sublattice, $A$ or $B$,  respectively, and $\vec{M}_2^{\alpha}$  is the total magnetization of the iron ions belonging to the
sublattice, $A$ or $B$, respectively. 
$\mathcal{J}_{ij} = P_i[x] P_j[x] J_{ij}$, $\mathcal{D}_{ij} = P_i[x] P_j[x] D_{ij}$, $\mathcal{K}_i = P_i[x]K_i$, and 
$\mathcal{H}_i = m_i P_i[x]H$ with $m_1=0.6$ and $m_2=1$. 
Let $\phi$  and $\theta$ the canting angles of $M_1$ and $M_2$ respectively (see figure \ref{fig0}).  
$z$ is the number of nearest neighbors. 

For small canting angles, disregarding constant terms,  the energy is

\begin{eqnarray}
\nonumber
E &=& 2 \mathcal{J}_{11}  \phi^2 + 2 \mathcal{J}_{22}  \theta^2  +   \mathcal{J}_{12}(\theta + \phi)^2  -  2 \mathcal{D}_{11}  \phi -  2 \mathcal{D}_{22}  \theta + \\  
  & + & 2 \mathcal{D}_{12} (\theta + \phi) +  2 \mathcal{K}_1 \phi^2 + 2 \mathcal{K}_2 \theta^2 - 2 \mathcal{H}_1 \phi - 2 \mathcal{H}_2 \theta.  
\end{eqnarray}
The minimum energy configuration is obtained from 
\begin{eqnarray}
\nonumber
\frac{\partial E}{\partial \theta} & = & 4 \mathcal{J}_{22}  \theta  +  2 \mathcal{J}_{12}(\theta + \phi)  -  2 \mathcal{D}_{22}  +  \mathcal{D}_{12} + 4 \mathcal{K}_2 \theta 
+ \\\nonumber 
                                   & - & 2 \mathcal{H}_2  = 0 \\ \nonumber
\frac{\partial E}{\partial \phi} & = & 4 \mathcal{J}_{11}  \phi  +  2 \mathcal{J}_{12}(\theta + \phi)  -  2 \mathcal{D}_{11}  +  \mathcal{D}_{12} + 4 \mathcal{K}_1 \phi + \\ \nonumber
                                   & - & 2 \mathcal{H}_1  = 0. 
\end{eqnarray}
Solving these two equations for $\theta$ and $\phi$ we can obtain the magnetization per site as function of $x$ as:
\begin{equation}
\label{mx}
M_z(x) = \mu_{cr} P_1[x]\phi(x) +  \mu_{Fe} P_2[x]\theta(x).
\end{equation}
In order to compare with MC simulations we define the reduced magnetization $ m_z = M_z/\mu_{Fe}$. 
For $x=0$,  $P_2=1$ and $P_1=0$ we have 

\begin{equation}
m_z=\frac{D_{22} + m_2 H}{2(J_{22} +  K_2)} m_2,
\end{equation}

\noindent  and for
$x=1$, $P_2=0$ and $P_1=1$, so

\begin{equation}
m_z=\frac{D_{11} + m_1 H }{2(J_{11}+ K_1)} m_1.
\end{equation}

\noindent These are the zero temperature --weak-- magnetization for the pure compounds, LuFeO$_3$ and LuCrO$_3$, 
respectively. 

\subsection{Experiments}

LuFe$_{1-x}$Cr$_x$O$_3$ ($x=0.15, 0.5$ and $0.85$) samples were prepared in polycrystalline form by a wet chemical method. 
A very reactive precursor was prepared starting from an aqueous solution of the metal ions and citric acid. 
Stoichiometric amounts of analytical grade Lu$_2$O$_3$, Fe(NO$_3$)$_3$ $\cdot$9H$_2$O and Cr(NO$_3$)$_3$ $\cdot$9H$_2$O were 
dissolved in citric acid and some drops of concentrated HNO$_3$, to facilitate the dissolution of Lu$_2$O$_3$. 
The citrate solution was  slowly evaporated, leading to an organic resin that contained a homogeneous distribution of the involved cations. 
This resin was dried at 120 $^o$C and then decomposed at 600 $^o$C for 12 h in air, with the aim of eliminate the 
organic matter. This treatment produced homogeneous and very reactive precursor materials that were finally 
treated at 1050 $^o$C in air for 12 h. LuFe$_{1-x}$Cr$_x$O$_3$ compounds were obtained as orange, well-crystallized powders 
as shown in Ref.[\onlinecite{Pomiro2016}].
The magnetic measurements were performed using a commercial MPMS-5S superconducting quantum interference 
device magnetometer, on powdered samples, from 5 to 400K, and for the 300 to 800K measurements the VSM option 
was used in the same MPMS.

\section{Results}
\label{sec:result}

\subsection{Antiferromagnetic ordering temperature}
\label{Tneel}
The analysis of the solid solution N\'eel temperature $T_N(x)$ of LuFe$_{1-x}$Cr$_x$O$_3$  as a function of the Cr 
concentration allowed us to estimate the coupling constants of the model as follows.

The critical temperature  obtained from MC in our model for $x=0$ (LuFeO$_3$) is $\frac{T_N}{J_{22}} = 1.44$. Considering 
that the measured N\'eel temperature\cite{Yuan2012}   is  $T_N = 628$ K, then the value for the superexchange interaction between 
the Fe$^{+3}$ ions that reproduces the experimental result in our model is: $J_{22} = 436$ K. 
Similarly, for $x=1$ (LuCrO$_3$),  $T_{N} = 115$ K\cite{Sahu2007}  and then $J_{11} = 79.8$ K. 
In our analysis the value of $J_{12}$ is a fitting parameter and it will be extracted from the approach of MC simulations 
and the experimental results.  Namely, we choose the value of   $J_{12}$ which provides an MC curve $T_N(x)$  that minimized 
the sum of the mean square deviations respect to the available experimental results.

Previous estimations of the solid  solution N\'eel temperature $T_N(x)$ in  this kind of compounds were based on 
 mean field approximations, in which  DM interactions were neglected\cite{Dasari12EPL,Hashimoto63JPSJ}.  For instance, 
Dasari et al.\cite{Dasari12EPL} obtained

\begin{equation}
\label{eq-dasari1}
T_N = \frac{z}{3}\left(\sum_{i=1,j=1}^2 J_{ij}^2 P_i[x]^2P_j[x]^2 \right)^{\frac{1}{2}},
\end{equation}
\noindent where $z$ is the number of nearest neighbors, $J_{ij}$ and $ P_i[x]$,  with $i,j=1,2$  where already defined 
in Sections \ref{MC} and \ref{EM}.
In a cubic lattice $z=6$,  so for the pure compounds ($x=0$ or $x=1$) $T_{N_i} = \frac{z}{3} J_{ii}$ and therefore

\begin{equation}
\label{eq-dasari2}
T_N = \sqrt{T_{N_1}^2 P_1[x]^4 + 8 J_{12}^2 P_1[x]^2 P_2[x]^2 + T_{N_2}^2 P_2[x]^4}.
\end{equation}
In a different mean field approximation Hashimoto~\cite{Hashimoto63JPSJ} obtained the expression 
\begin{widetext}
\begin{equation}
\label{eq-hashi-1}
T_N = \frac{1}{2} \left[P_1[x]T_{N_1} - P_2[x]T_{N_2}  + \sqrt{ (P_1[x]T_{N_1} +P_2[x] T_{N_2})^2 + 4P_1[x]P_2[x](  \frac{4 z^2}{9}J_{12}^2 - T_{N_1} T_{N_2})} \right].
\end{equation}
\end{widetext}

\begin{figure}
\begin{center}
\includegraphics[scale=0.35, angle=-90]{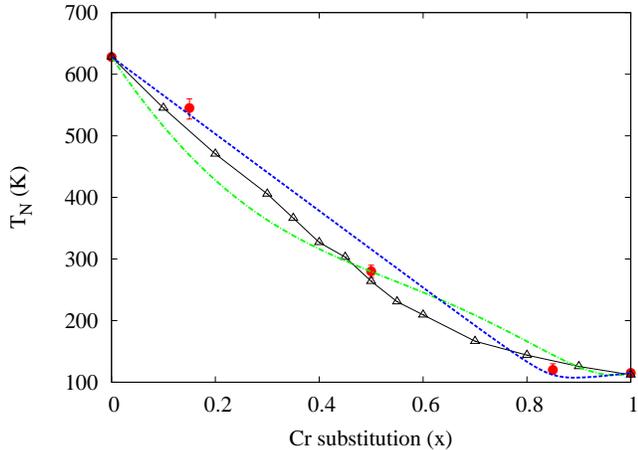}
\caption{\label{fig1} (Color online)  
N\'eel temperature as function of the fraction of Cr. Full red circles correspond to the experiments with  LuFe$_{1-x}$Cr$_x$O$_3$,
and open triangles to MC simulations.  The lines correspond to fits of the experiments using
Eq.(\ref{eq-dasari2}) (blue dashed lines)  and Eq.(\ref{eq-hashi-1}) (green dot dashed lines), giving $J_{12}= 162$ K and $J_{12}= 11$ K, respectively. 
The parameter values of the MC simulation were  $J_{22} = 436$ K, $J_{11} = 79.8$ K, and $J_{12} = 106$ K. 
}
\end{center}
\end{figure}

The dependence of the N\'eel temperature on the Cr content obtained from experiments in polycrystals is shown in
Fig. \ref{fig1}. The values $x=0$\cite{Yuan2012}, $x=1$\cite{Sahu2007} 
were taken from  the literature, 
and $x=0.15, 0.5$ and $0.85$ were sinthetized in our experiments. In this figure we also compare the best fittings of 
the experimental results obtained from the MC simulations and 
using Eqs. (\ref{eq-dasari2}) and (\ref{eq-hashi-1}). 

From Hashimoto   and Dasari  expressions very different values of $J_{12}$ are obtained, $11$ K and $162$ K respectively. The value derived
from the MC simulations  is $J_{12}=106$ K,  which is in between the values obtained from Eqs.(\ref{eq-hashi-1}) and(\ref{eq-dasari2}).
Considering that the value of the exchange integral $J_{FeCr}=J_{12}/(2 S_{Cr} S_{Fe})$ where $S_{Cr}^2 = S_{Cr}(S_{Cr}+1)$ 
and $S_{Fe}^2 = S_{Fe}(S_{Fe}+1)$ we obtain  $J_{FeCr}= 28,3$ K,  $J_{FeCr}= 1.9$ K  and $J_{FeCr} = 9.25$ K
from Eq. (\ref{eq-hashi-1}),  Eq. (\ref{eq-dasari2}) and MC simulations, respectively.
The value of $J_{FeCr}$ reported by Dasari et al [\onlinecite{Dasari12EPL}] for  YFe$_{1-x}$Cr$_x$O$_3$
($J_{FeCr}= 24.3$ K) is comparably to the value we have found for LuFe$_{1-x}$Cr$_x$O$_3$ using
the same equation.
The value reported by Kadomtseva et al. for YFe$_{1-x}$Cr$_x$O$_3$ monocrystals using Eq. (\ref{eq-hashi-1})
($J_{FeCr}=6,64$ K)  is considerably lower than the value reported by Dasari et al. in the
same compound.   

\begin{figure}
\begin{center}
\includegraphics[scale=0.35, angle=-90]{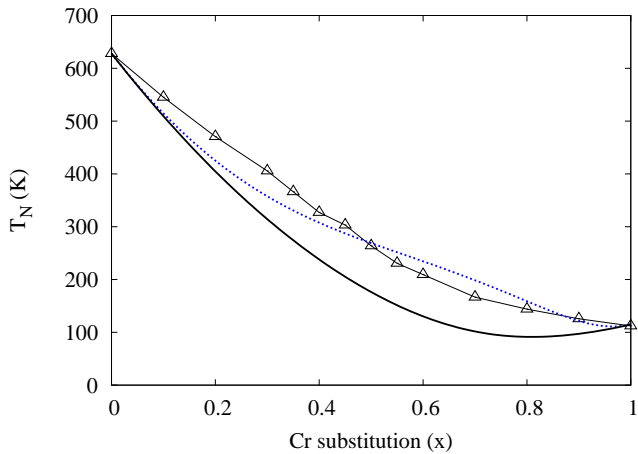}
\caption{\label{fig1b} (Color online)  
N\'eel temperature as function of the fraction of Cr for LuFe$_{1-x}$Cr$_x$O$_3$. Open triangles correspond to MC simulations 
and the blue dotted line corresponds to a fit of the MF expression Eq. (\ref{eq-dasari2}). The parameter that results from the fit  
is $J_{12} = 306$ K. 
The full line corresponds to a plot of Eq.  (\ref{eq-dasari2})  using the value of MC simulations  ($J_{12}= 106)$.
}
\end{center}
\end{figure}

In order to test the mean field approximations,  we fitted the MC results with the corresponding expressions 
(\ref{eq-dasari2}) and (\ref{eq-hashi-1}). 
 In figure \ref{fig1b} we show a fit of the N\'eel temperatures 
obtained from  MC simulation using Eq.(\ref{eq-dasari2}).
From this fit we obtained $J_{12} = 306$ K which is considerably greater than the value used in MC simulations 
$ J_{12} = 106$ K. 
For comparison we included in Fig. \ref{fig1b}  a plot of the mean field expression
using the MC simulation value  ($J_{12} = 106$ K). One can observe that with this value Eq.(\ref{eq-dasari2})
clearly departs from the results of MC simulations.   
We concluded that a fit with expression (\ref{eq-dasari2}) always overestimates the value of the exchange interaction $J_{12}$. 
Similarly, fitting the MC results using Eq.(\ref{eq-hashi-1}) systematically underestimates $J_{12}$. The 
accuracy of  MF model is expected to be good in low and high Cr content; at intermediate concentrations the effect
of the distribution of the interactions is important. Then, a fit which takes into account all the concentration
range somehow biases  the value of the $J_{12}$ interaction.

\begin{figure}
\begin{center}
\includegraphics[scale=0.35, angle=-90]{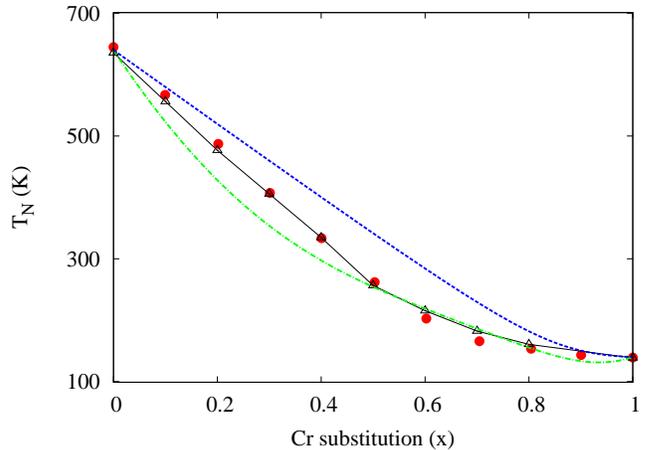}
\caption{\label{fig4} (Color online)  
N\'eel temperatures as function of the fraction of Cr for~\cite{Dasari12EPL}  YFe$_{1-x}$Cr$_x$0$_3$
(red circles) and MC simulations (open triangles). 
The blue dashed line corresponds to Eq.(\ref{eq-dasari2}) using the value of $J_{12}$ reported by Dasari et al [\onlinecite{Dasari12EPL}] 
and the green dot dashed line corresponds to Eq. (\ref{eq-hashi-1}) using the value of $J_{12}$ reported by Kadomtseva et al [\onlinecite{Kadomtseva77JETP}] 
}
\end{center}
\end{figure}

In figure \ref{fig4} we show experimental data for the critical temperature of  YFe$_{1-x}$Cr$_x$O$_3$ as a function of the
chromium content reported by Dasari et al. \cite{Dasari12EPL}. We also show the data obtained from
MC simulations with $J_{FeCr} = 9.25$ K tuned to get the best fit with the experimental
points. We also include a plot of the mean field expression derived by Dasari et al. \cite{Dasari12EPL} 
using the value  $J_{FeCr} = 24.0$ K which is the value reported by these authors. Finally, a plot of
Hashimoto's expression \cite{Hashimoto63JPSJ}  Eq.(\ref{eq-hashi-1}) using the value of $J_{FeCr} = 6.64$ K 
reported by Kadomtseva et al \cite{Kadomtseva77JETP} is also included. In this last work the samples 
studied were  single monocrystals  of the YFe$_{1-x}$Cr$_x$O$_3$ compound.
MC approach gives a very good agreement with the experiments in all the range of concentrations and
the value obtained for $J_{FeCr} = 9.25$ K is higher than the reported by Kadomtseva et al.  and lower
 than the reported by  Dasari et al. 
Finally, the values $J_{FeCr}$ obtained from MC  for both compounds YFe$_{1-x}$Cr$_x$O$_3$ and 
LuFe$_{1-x}$Cr$_x$O$_3$ are the same, indicating that the exchange integral is not substantially affected 
by the substitution of yttrium by lutetium.

\subsection{$T=0$ magnetization}

\begin{figure}
\begin{center}
\includegraphics[scale=0.35, angle=-90]{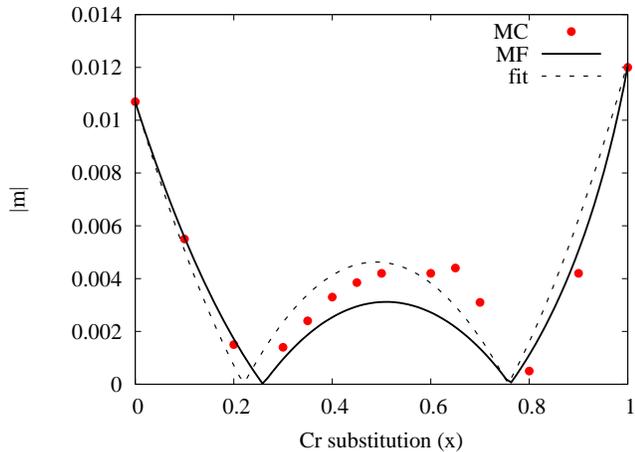}
\caption{\label{fig3} (Color online)  
Module of the canted magnetization as function of the Cr content at $T=0$ and zero applied magnetic
field. Monte Carlo simulations (MC): red circles. Full lines correspond to the 
effective model Eq. (\ref{mx}) using the same parameters as in the simulation. The dashed lines correspond to a fit
using Eq. (\ref{eq-fit}).  
}
\end{center}
\end{figure}

In Fig.\ref{fig3} we show the modulus of the canted magnetization in the z direction ($m_z$) at $T=0$ as 
function of the Cr content which is obtained from MC simulations in a ZFC process. We also plot the modulus 
of $m_z$  obtained  using  Eq. (\ref{mx})  with the physical constants used in MC simulations. 
We see a good agreement between MC and the effective model at low and high chromium contents where
the model is expected to work better. 
The local maximum at intermediate concentrations observed in MC simulations is related to a 
change in the sign of the magnetization.
Like in the mean field model at intermediate concentrations the effect of the distributions of the DM 
bonds is important, and for this reason in this concentration range the effective model departs from  
MC simulations, in fact the effective model takes into account only averaged values in the distribution 
of the DM interactions.
In addition, the effective canting due to the DM interactions can be approached using the
following expression for the magnetization as function of the chromium content
\begin{equation}
\label{eq-fit}
m_z = \frac{1}{2}[m_{22}P_1[x]^2 + 2 m_{12}J_{12}^2 P_1[x]P_2[x] + m_{22} P_2[x]^2]
\end{equation}
where $m_{ij}$ is the averaged canted magnetization contribution of a pair of spins interacting through the
DM interaction. Then, $m_{11} = 2 \alpha_{11} m_{1}$, $m_{12} =\alpha_{12} (m_{1} +  m_{2})$, 
and $m_{22} = 2 \alpha_{22} m_{2}$. Here $\alpha_{ij}$ are the average canting angles between 
 ions of type $i$ and $j$ (assuming low angles).
According to Eq. (\ref{mx}), $\alpha_{ij} \simeq \frac{D_{ij}}{2J_{ij}}$ for $i=j$, and $\alpha_{12}$ 
is an effective parameter to be fitted. One can estimate as  $\alpha_{12}=\frac{D_{12}}{2J_{21}}$
which turn in $\alpha_{12}=-0.0345$. The fixed parameters are $\alpha_{11}=0.0203$
and $\alpha_{22}=0.0107$ . Fitting Eq. (\ref{eq-fit}) to the MC data we obtain $\alpha_{12}=-0.0258$, 
showing the consistency of Eq.(\ref{eq-fit}).   Moreover, the negative magnetization can be understood 
from Eq.(\ref{eq-fit}) as an effect of the negative sign of the DM interactions between Fe and Cr ions, 
which favors  the canting of both ions in the negative z direction. Since Fe-Cr pairs are the majority 
at intermediate values of the Cr concentration $x$, the term in Eq.(\ref{eq-fit}) associated with 
$m _{12}<0$  is dominant and $m_z$ becomes negative.

This local maximum has been reported in experiments carried out in single crystals of the YFe$_{1-x}$Cr$_x$O$_3$ 
compounds~\cite{Kadomtseva77JETP} where the low temperature magnetization is measured as function 
of the chromium content.

\subsection{Magnetization reversal}

\begin{figure}
\begin{center}
\includegraphics[scale=0.35, angle=-90]{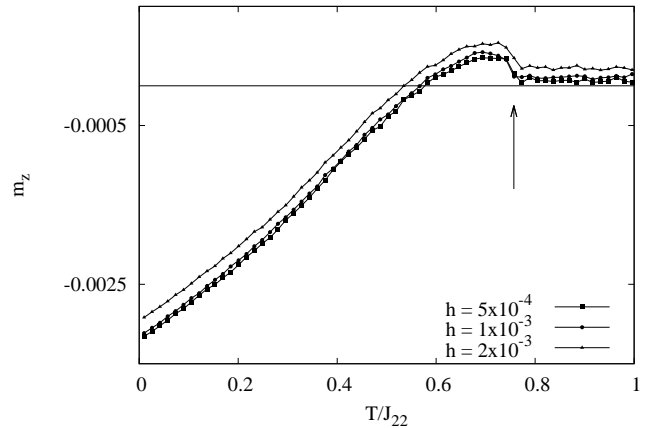}
\caption{\label{fig6}  
Field cooling curves under different applied magnetic fields in the $z$ direction for 
$x=0.4$.  
The arrow indicates the ordering temperature obtained from the peak of the
susceptibility. The applied field is in units of $J_{22}$ i.e. $h=H/J_{22}$. 
}
\end{center}
\end{figure}

In Fig. \ref{fig6} we show the  $m_z$  component of the total magnetization as function 
of the temperature  for $x=0.4$ chromium content obtained by MC simulations using the parameters 
of the LuFe$_{1-x}$Cr$_x$O$_3$ compound already obtained in section \ref{Tneel}.  
The cooling is performed under three different applied fields in the $z$ direction; $h=H/J_{22}=5\times10^{-4},\, 1\times10^{-3}$,
and $2\times10^{-3}$. The arrow indicates the N\'eel temperature for this composition. 
We can observe reversal in the magnetization for this composition for the three  applied fields. 
The magnetization increases with the applied field at low temperatures, although  the compensation temperature 
appears to be almost independent of $h$, at least in the small range of values.
In this case, $x=0.4$, the compensation temperature is clearly smaller than the N\'eel temperature. 
We do not observe magnetization  reversal for $x=0.5$ and in fact in the composition range
$x > 0.4  $ the magnetization reversal is unstable.
However, the shape of the curve obtained for $x=0.4$ qualitatively  reproduces the curves
reported in the experiments  for yttrium \cite{Mao11APL} (YFe$_{0.5}$Cr$_{0.5}$O$_3$) and 
lutetium~\cite{Pomiro2016} (LuFe$_{0.5}$Cr$_{0.5}$O$_3$) compounds.  Moreover, the ratio between 
the compensation and N\'eel temperatures obtained in our simulation $T_{comp}/T_N = 0.76 $ compares 
well with the experimental value\cite{Pomiro2016} $T_{comp}/T_N = 0.83$.

\begin{figure}
\begin{center}
\includegraphics[scale=0.35, angle=-90]{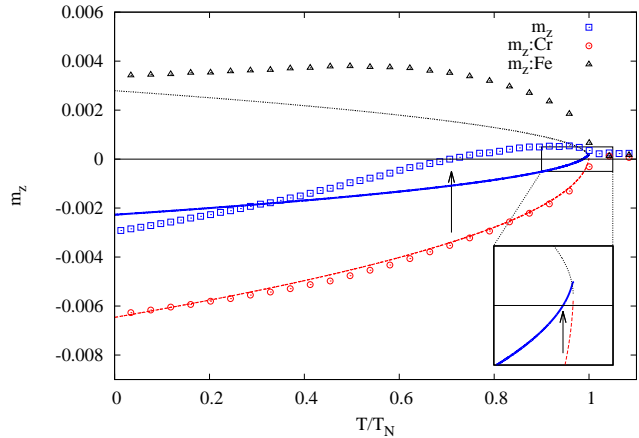}
\caption{\label{fig7} (Color online)  
Field cooling curves with an applied fields in the $z$ direction for  
$x=0.4$. The applied field ($h=0.002$) is the same for all the curves.  
The symbols with lines correspond to MC simulations and lines to MF results.
MC simulations: chromium magnetization (black triangles), iron magnetization (red circles) and total magnetization (blue squares). 
MF calculations: chromium magnetization (black dotted line), iron magnetization (red dashed line) and total magnetization (blue continuous line). 
The arrows indicate  the  compensation temperatures in both cases, and the inset is a zoom of MF curves close to the ordering temperature. 
}
\end{center}
\end{figure}

In Fig. \ref{fig7} we show FC magnetization curves with $h=2\times 10^{-3} $ for $x=0.4$,
and curves obtained through a mean field approach. 
Here we have measured in MC simulations separately the temperature dependence of 
the total magnetization of the Fe$^{3+}$ ions and that of the  Cr$^ {3+}$ ions.   
We can see that below the N\'eel temperature the  magnetization due to the Fe$^{3+}$ ions aligns 
in  the direction of the magnetic field, while the magnetization due to the Cr$^{3+}$ ions 
is opposite to the field. In this way, when the field breaks the inversion symmetry along the $z$ axis 
the Zeeman energy is reduced  due to the coupling of the larger magnetic moments of the Fe$^{3+}$ 
ions. The different temperatures dependencies in the magnetization of Fe$^{3+}$ and Cr$^{3+}$
ions turns into the magnetization reversal. For lower compositions ($x \leq 0.3$) the Fe$^{3+}$ ions 
are also aligned in the direction of the applied field but magnetization reversal is not observed 
because the  contribution to the magnetization of Fe$^{3+}$ is dominant. 
 
In this figure we also show mean field curves which are obtained through Eq. (\ref{eq-mf}) using
a molecular field approximation for the dependence of the sublattice magnetization on the temperature.
This approximation agrees very well with MC results for the sublattice magnetization. 
In the calculation of the mean field curves showed in Fig. Fig. \ref{fig7} we used the same  
parameters than in MC results. 
These curves reproduce the features observed in MC results. However, 
in this case the compensation temperature (see inset) is much closer to the N\'eel temperature.    
From an analysis of the different energy term contributions, Eq. (\ref{eq-mf}), we observed
that close to the N\'eel temperature the Zeeman term is the most important hence the coupling
with the field at high temperatures rules the magnetization process and induces the symmetry breaking. 
In the  lower temperature range DM interactions prevail and convey the reversal of the magnetization.

\section{Discussion}
\label{title = {"El paper experimental"}, sec:discussion}
Monte Carlo simulations using the  proposed microscopic classical model reproduce the whole phenomenology  
of  both LuFe$_{1-x}$Cr$_x$O$_3$ and YFe$_{1-x}$Cr$_x$O$_3$ compounds as the chromium content is varied.
From these simulations it turns out that the superexchange interaction  between Cr$^{3+}$ and Fe$^{3+}$ ions is 
lower than the super-exchange interaction between Fe$^{3+}$ and Fe$^{3+}$, and greater than the super-exchange between 
Cr$^{3+}$ and Fe$^{3+}$ ions i.e.  $J_{22} > J_{12} > J_{11}$. 
From the fit of our experimental results with different mean field expressions,  Eqs. (\ref{eq-dasari2}) and (\ref{eq-hashi-1})
we obtained $J_{12} = 162$ K and $J_{12} = 11$ K, respectively.  

These results show a big dispersion depending of the expression used to fit the experiments. In particular,  
the value obtained from MC simulations,  $J_{12}=106$ K,   is in between this two values.  
The values of $J_{12}$ available in the literature for Y perovskites YFe$_{1-x}$Cr$_x$O$_3$, also show an important  
dispersion. For instance,  $J_{12} = 139$ K, when  Eq. (\ref{eq-dasari2}) is used in polycrystals~\cite{Dasari12EPL} and  
$J_{12} = 38$ K has been reported in single crystals~\cite{Kadomtseva77JETP}  using Eq. (\ref{eq-hashi-1}).  
Such large sensitivity to the details of the particular mean field approximation is not surprising in a solid solution, 
where the interplay between thermal fluctuations  and the inherent disorder of the solution is expected to be very 
relevant to determine thermal properties. Consistently, the experimental results are better described by the MC 
simulations than by the MF expressions. Hence, we expect our  estimation of  $J_{12}$ to be more reliable than the 
previous ones. In addition, our results suggest that the exchange constant (and therefore the general behavior)  
is not substantially affected  by the substitution of yttrium by lutecium.

The zero temperature magnetization obtained from MC simulations in a ZFC process, which is due to the canting 
of the  AFM spins  in the $z$ directions, is well approached by an effective coarse grain model in the range 
of low and high chromium contents as expected.
A bump in the magnetization is observed in MC simulations at intermediate concentrations which is a signature of 
the magnetization reversal. This bump is also observed in the coarse grain approach but is less pronounced. 
The difference  between MC simulations and the effective  model at intermediate concentrations is expected since 
the coarse grain approach does not take into account information on the distribution of the ions in the lattice 
which is important at intermediate Cr concentrations. 

Magnetization reversal is observed at intermediate chromium contents $x=0.4$  in a ZFC process depending on 
the value of $D_{12} (\simeq 1.7 \times 10^{-2}J_{22})$. 
When the field is increased above a certain  threshold MR disappear, and the magnetization points in the 
direction of the applied field in whole temperature range.
The presence of  magnetization reversal is very sensitive to the value of the DM interaction between  Cr$^{3+}$ 
and Fe$^{3+}$ ions and also to the value of superexchange $J_{12}$ interaction.
We do not observe magnetization reversal in MC simulations at $x=0.5$ such as is observed in
experiments. This could be due to size effects which are particularly important in systems that includes 
disorder. In fact, the fields we used to obtain the ZFC curves (e.g. $h=0.001$ correspond to $B \simeq 0.13$ T) are 
much greater that the ones used in experiments (e.g. $B \sim 0.01$ T). A reduction of the fields
in MC is only possible in larger systems.

Summarizing, Monte Carlo simulations based in a Heisenberg microscopic classical model reproduce 
the critical temperatures observed in experiments. 
Besides this is a classical model, MC fit can provide a better estimation of $J_{12}$ since in
this model the random occupation of the Cr$^{3+}$ and Fe$^{3+}$ ions is taken into account.
Regarding the phenomena of magnetization reversal, we found it for appropriated values of the superexchange 
and the  Dzyaloshinskii-Moriya interactions at intermediate Cr concentrations. However, the mechanism
for the appearance is subtle and further investigations are needed to shed light on this point.

\acknowledgements

This work was partially supported by CONICET  through grant PIP 2012-11220110100213 and PIP 2013-11220120100360, 
SeCyT--Universidad Nacional de C\'ordoba (Argentina),  FONCyT and a CONICET-CNRS cooperation program. 
F. P. thanks CONICET for a fellowship. A. M. gratefully acknowledges a collaboration project between CNRS and 
CONICET (PCB I-2014). 
This work used Mendieta Cluster from CCAD-UNC, which is part of SNCAD-MinCyT, Argentina.


\begin{thebibliography}{28}%
\makeatletter
\providecommand \@ifxundefined [1]{%
 \@ifx{#1\undefined}
}%
\providecommand \@ifnum [1]{%
 \ifnum #1\expandafter \@firstoftwo
 \else \expandafter \@secondoftwo
 \fi
}%
\providecommand \@ifx [1]{%
 \ifx #1\expandafter \@firstoftwo
 \else \expandafter \@secondoftwo
 \fi
}%
\providecommand \natexlab [1]{#1}%
\providecommand \enquote  [1]{``#1''}%
\providecommand \bibnamefont  [1]{#1}%
\providecommand \bibfnamefont [1]{#1}%
\providecommand \citenamefont [1]{#1}%
\providecommand \href@noop [0]{\@secondoftwo}%
\providecommand \href [0]{\begingroup \@sanitize@url \@href}%
\providecommand \@href[1]{\@@startlink{#1}\@@href}%
\providecommand \@@href[1]{\endgroup#1\@@endlink}%
\providecommand \@sanitize@url [0]{\catcode `\\12\catcode `\$12\catcode
  `\&12\catcode `\#12\catcode `\^12\catcode `\_12\catcode `\%12\relax}%
\providecommand \@@startlink[1]{}%
\providecommand \@@endlink[0]{}%
\providecommand \url  [0]{\begingroup\@sanitize@url \@url }%
\providecommand \@url [1]{\endgroup\@href {#1}{\urlprefix }}%
\providecommand \urlprefix  [0]{URL }%
\providecommand \Eprint [0]{\href }%
\providecommand \doibase [0]{http://dx.doi.org/}%
\providecommand \selectlanguage [0]{\@gobble}%
\providecommand \bibinfo  [0]{\@secondoftwo}%
\providecommand \bibfield  [0]{\@secondoftwo}%
\providecommand \translation [1]{[#1]}%
\providecommand \BibitemOpen [0]{}%
\providecommand \bibitemStop [0]{}%
\providecommand \bibitemNoStop [0]{.\EOS\space}%
\providecommand \EOS [0]{\spacefactor3000\relax}%
\providecommand \BibitemShut  [1]{\csname bibitem#1\endcsname}%
\let\auto@bib@innerbib\@empty
\bibitem [{\citenamefont {Kadomtseva}\ \emph {et~al.}(1977)\citenamefont
  {Kadomtseva}, \citenamefont {Moskvin}, \citenamefont {Bostrem}, \citenamefont
  {Wanklyn},\ and\ \citenamefont {Khafizova}}]{Kadomtseva77JETP}%
  \BibitemOpen
  \bibfield  {author} {\bibinfo {author} {\bibfnamefont {A.~M.}\ \bibnamefont
  {Kadomtseva}}, \bibinfo {author} {\bibfnamefont {A.~S.}\ \bibnamefont
  {Moskvin}}, \bibinfo {author} {\bibfnamefont {I.~G.}\ \bibnamefont
  {Bostrem}}, \bibinfo {author} {\bibfnamefont {B.~M.}\ \bibnamefont
  {Wanklyn}}, \ and\ \bibinfo {author} {\bibfnamefont {N.~A.}\ \bibnamefont
  {Khafizova}},\ }\href@noop {} {\bibfield  {journal} {\bibinfo  {journal}
  {Sov. Phys. JETP}\ }\textbf {\bibinfo {volume} {45}},\ \bibinfo {pages}
  {1202} (\bibinfo {year} {1977})}\BibitemShut {NoStop}%
\bibitem [{\citenamefont {Yoshii}\ and\ \citenamefont
  {Nakamura}(2000)}]{Yoshii00JSSC}%
  \BibitemOpen
  \bibfield  {author} {\bibinfo {author} {\bibfnamefont {K.}~\bibnamefont
  {Yoshii}}\ and\ \bibinfo {author} {\bibfnamefont {A.}~\bibnamefont
  {Nakamura}},\ }\href@noop {} {\bibfield  {journal} {\bibinfo  {journal}
  {Journal of Solid State Chemistry}\ }\textbf {\bibinfo {volume} {155}},\
  \bibinfo {pages} {447 } (\bibinfo {year} {2000})}\BibitemShut {NoStop}%
\bibitem [{\citenamefont {Yoshii}(2001)}]{Yoshii01aJSSC}%
  \BibitemOpen
  \bibfield  {author} {\bibinfo {author} {\bibfnamefont {K.}~\bibnamefont
  {Yoshii}},\ }\href@noop {} {\bibfield  {journal} {\bibinfo  {journal}
  {Journal of Solid State Chemistry}\ }\textbf {\bibinfo {volume} {159}},\
  \bibinfo {pages} {204 } (\bibinfo {year} {2001})}\BibitemShut {NoStop}%
\bibitem [{\citenamefont {Yoshii}\ \emph {et~al.}(2001)\citenamefont {Yoshii},
  \citenamefont {Nakamura}, \citenamefont {Ishii},\ and\ \citenamefont
  {Morii}}]{Yoshii01bJSSC}%
  \BibitemOpen
  \bibfield  {author} {\bibinfo {author} {\bibfnamefont {K.}~\bibnamefont
  {Yoshii}}, \bibinfo {author} {\bibfnamefont {A.}~\bibnamefont {Nakamura}},
  \bibinfo {author} {\bibfnamefont {Y.}~\bibnamefont {Ishii}}, \ and\ \bibinfo
  {author} {\bibfnamefont {Y.}~\bibnamefont {Morii}},\ }\href@noop {}
  {\bibfield  {journal} {\bibinfo  {journal} {Journal of Solid State
  Chemistry}\ }\textbf {\bibinfo {volume} {162}},\ \bibinfo {pages} {84 }
  (\bibinfo {year} {2001})}\BibitemShut {NoStop}%
\bibitem [{\citenamefont {Mao}\ \emph {et~al.}(2011{\natexlab{a}})\citenamefont
  {Mao}, \citenamefont {Sui}, \citenamefont {Zhang}, \citenamefont {Su},
  \citenamefont {Wang}, \citenamefont {Liu}, \citenamefont {Wang},
  \citenamefont {Zhu}, \citenamefont {Wang}, \citenamefont {Liu},\ and\
  \citenamefont {Tang}}]{Mao11APL}%
  \BibitemOpen
  \bibfield  {author} {\bibinfo {author} {\bibfnamefont {J.}~\bibnamefont
  {Mao}}, \bibinfo {author} {\bibfnamefont {Y.}~\bibnamefont {Sui}}, \bibinfo
  {author} {\bibfnamefont {X.}~\bibnamefont {Zhang}}, \bibinfo {author}
  {\bibfnamefont {Y.}~\bibnamefont {Su}}, \bibinfo {author} {\bibfnamefont
  {X.}~\bibnamefont {Wang}}, \bibinfo {author} {\bibfnamefont {Z.}~\bibnamefont
  {Liu}}, \bibinfo {author} {\bibfnamefont {Y.}~\bibnamefont {Wang}}, \bibinfo
  {author} {\bibfnamefont {R.}~\bibnamefont {Zhu}}, \bibinfo {author}
  {\bibfnamefont {Y.}~\bibnamefont {Wang}}, \bibinfo {author} {\bibfnamefont
  {W.}~\bibnamefont {Liu}}, \ and\ \bibinfo {author} {\bibfnamefont
  {J.}~\bibnamefont {Tang}},\ }\href@noop {} {\bibfield  {journal} {\bibinfo
  {journal} {Applied Physics Letters}\ }\textbf {\bibinfo {volume} {98}},\
  \bibinfo {eid} {192510} (\bibinfo {year} {2011}{\natexlab{a}})}\BibitemShut
  {NoStop}%
\bibitem [{\citenamefont {Dasari}\ \emph {et~al.}(2012)\citenamefont {Dasari},
  \citenamefont {Mandal}, \citenamefont {Sundaresan},\ and\ \citenamefont
  {Vidhyadhiraja}}]{Dasari12EPL}%
  \BibitemOpen
  \bibfield  {author} {\bibinfo {author} {\bibfnamefont {N.}~\bibnamefont
  {Dasari}}, \bibinfo {author} {\bibfnamefont {P.}~\bibnamefont {Mandal}},
  \bibinfo {author} {\bibfnamefont {A.}~\bibnamefont {Sundaresan}}, \ and\
  \bibinfo {author} {\bibfnamefont {N.~S.}\ \bibnamefont {Vidhyadhiraja}},\
  }\href@noop {} {\bibfield  {journal} {\bibinfo  {journal} {EPL (Europhysics
  Letters)}\ }\textbf {\bibinfo {volume} {99}},\ \bibinfo {pages} {17008}
  (\bibinfo {year} {2012})}\BibitemShut {NoStop}%
\bibitem [{\citenamefont {Mandal}\ \emph {et~al.}(2013)\citenamefont {Mandal},
  \citenamefont {Serrao}, \citenamefont {Suard}, \citenamefont {Caignaert},
  \citenamefont {Raveau}, \citenamefont {Sundaresan},\ and\ \citenamefont
  {Rao}}]{Mandal13JSSC}%
  \BibitemOpen
  \bibfield  {author} {\bibinfo {author} {\bibfnamefont {P.}~\bibnamefont
  {Mandal}}, \bibinfo {author} {\bibfnamefont {C.}~\bibnamefont {Serrao}},
  \bibinfo {author} {\bibfnamefont {E.}~\bibnamefont {Suard}}, \bibinfo
  {author} {\bibfnamefont {V.}~\bibnamefont {Caignaert}}, \bibinfo {author}
  {\bibfnamefont {B.}~\bibnamefont {Raveau}}, \bibinfo {author} {\bibfnamefont
  {A.}~\bibnamefont {Sundaresan}}, \ and\ \bibinfo {author} {\bibfnamefont
  {C.}~\bibnamefont {Rao}},\ }\href@noop {} {\bibfield  {journal} {\bibinfo
  {journal} {Journal of Solid State Chemistry}\ }\textbf {\bibinfo {volume}
  {197}},\ \bibinfo {pages} {408 } (\bibinfo {year} {2013})}\BibitemShut
  {NoStop}%
\bibitem [{\citenamefont {Ren}\ \emph {et~al.}(1998)\citenamefont {Ren},
  \citenamefont {Palstra}, \citenamefont {Khomskii}, \citenamefont {Pellegrin},
  \citenamefont {Nugroho}, \citenamefont {Menovsky},\ and\ \citenamefont
  {Sawatzky}}]{Ren1998}%
  \BibitemOpen
  \bibfield  {author} {\bibinfo {author} {\bibfnamefont {Y.}~\bibnamefont
  {Ren}}, \bibinfo {author} {\bibfnamefont {T.~T.~M.}\ \bibnamefont {Palstra}},
  \bibinfo {author} {\bibfnamefont {D.~I.}\ \bibnamefont {Khomskii}}, \bibinfo
  {author} {\bibfnamefont {E.}~\bibnamefont {Pellegrin}}, \bibinfo {author}
  {\bibfnamefont {A.~A.}\ \bibnamefont {Nugroho}}, \bibinfo {author}
  {\bibfnamefont {A.~A.}\ \bibnamefont {Menovsky}}, \ and\ \bibinfo {author}
  {\bibfnamefont {G.~A.}\ \bibnamefont {Sawatzky}},\ }\href {\doibase
  10.1038/24802} {\bibfield  {journal} {\bibinfo  {journal} {Nature}\ }\textbf
  {\bibinfo {volume} {396}},\ \bibinfo {pages} {441} (\bibinfo {year}
  {1998})}\BibitemShut {NoStop}%
\bibitem [{\citenamefont {Treves}(1962)}]{Treves62PR}%
  \BibitemOpen
  \bibfield  {author} {\bibinfo {author} {\bibfnamefont {D.}~\bibnamefont
  {Treves}},\ }\href@noop {} {\bibfield  {journal} {\bibinfo  {journal} {Phys.
  Rev.}\ }\textbf {\bibinfo {volume} {125}},\ \bibinfo {pages} {1843} (\bibinfo
  {year} {1962})}\BibitemShut {NoStop}%
\bibitem [{\citenamefont {Moriya}(1960)}]{Moriya1960}%
  \BibitemOpen
  \bibfield  {author} {\bibinfo {author} {\bibfnamefont {T.}~\bibnamefont
  {Moriya}},\ }\href@noop {} {\bibfield  {journal} {\bibinfo  {journal}
  {Physical Review Letters}\ }\textbf {\bibinfo {volume} {4}},\ \bibinfo
  {pages} {228} (\bibinfo {year} {1960})}\BibitemShut {NoStop}%
\bibitem [{\citenamefont {W.}\ and\ \citenamefont {A.}(1953)}]{Gorter1953}%
  \BibitemOpen
  \bibfield  {author} {\bibinfo {author} {\bibfnamefont {G.~E.}\ \bibnamefont
  {W.}}\ and\ \bibinfo {author} {\bibfnamefont {S.~J.}\ \bibnamefont {A.}},\
  }\href@noop {} {\bibfield  {journal} {\bibinfo  {journal} {Physical Review}\
  }\textbf {\bibinfo {volume} {90}},\ \bibinfo {pages} {487} (\bibinfo {year}
  {1953})}\BibitemShut {NoStop}%
\bibitem [{\citenamefont {N.}\ \emph {et~al.}(1960)\citenamefont {N.},
  \citenamefont {K.},\ and\ \citenamefont {G.}}]{Menyuk1960}%
  \BibitemOpen
  \bibfield  {author} {\bibinfo {author} {\bibfnamefont {M.}~\bibnamefont
  {N.}}, \bibinfo {author} {\bibfnamefont {D.}~\bibnamefont {K.}}, \ and\
  \bibinfo {author} {\bibfnamefont {W.~D.}\ \bibnamefont {G.}},\ }\href@noop {}
  {\bibfield  {journal} {\bibinfo  {journal} {Physical Review Letters}\
  }\textbf {\bibinfo {volume} {4}},\ \bibinfo {pages} {119} (\bibinfo {year}
  {1960})}\BibitemShut {NoStop}%
\bibitem [{\citenamefont {R.}(1958)}]{Pauthenet1958}%
  \BibitemOpen
  \bibfield  {author} {\bibinfo {author} {\bibfnamefont {P.}~\bibnamefont
  {R.}},\ }\href@noop {} {\bibfield  {journal} {\bibinfo  {journal} {J. Appl.
  Phys}\ }\textbf {\bibinfo {volume} {29}},\ \bibinfo {pages} {253} (\bibinfo
  {year} {1958})}\BibitemShut {NoStop}%
\bibitem [{\citenamefont {Ren}\ \emph {et~al.}(2000)\citenamefont {Ren},
  \citenamefont {Palstra}, \citenamefont {Khomskii}, \citenamefont {Nugroho},
  \citenamefont {Menovsky},\ and\ \citenamefont {Sawatzky}}]{Ren2000}%
  \BibitemOpen
  \bibfield  {author} {\bibinfo {author} {\bibfnamefont {Y.}~\bibnamefont
  {Ren}}, \bibinfo {author} {\bibfnamefont {T.~T.~M.}\ \bibnamefont {Palstra}},
  \bibinfo {author} {\bibfnamefont {D.~I.}\ \bibnamefont {Khomskii}}, \bibinfo
  {author} {\bibfnamefont {A.~A.}\ \bibnamefont {Nugroho}}, \bibinfo {author}
  {\bibfnamefont {A.~A.}\ \bibnamefont {Menovsky}}, \ and\ \bibinfo {author}
  {\bibfnamefont {G.~A.}\ \bibnamefont {Sawatzky}},\ }\href@noop {} {\bibfield
  {journal} {\bibinfo  {journal} {Physical Review B}\ }\textbf {\bibinfo
  {volume} {62}},\ \bibinfo {pages} {6577} (\bibinfo {year}
  {2000})}\BibitemShut {NoStop}%
\bibitem [{\citenamefont {Mao}\ \emph {et~al.}(2011{\natexlab{b}})\citenamefont
  {Mao}, \citenamefont {Sui}, \citenamefont {Zhang}, \citenamefont {Su},
  \citenamefont {Wang}, \citenamefont {Liu}, \citenamefont {Wang},
  \citenamefont {Zhu}, \citenamefont {Wang}, \citenamefont {Liu},\ and\
  \citenamefont {Tang}}]{Mao2011}%
  \BibitemOpen
  \bibfield  {author} {\bibinfo {author} {\bibfnamefont {J.}~\bibnamefont
  {Mao}}, \bibinfo {author} {\bibfnamefont {Y.}~\bibnamefont {Sui}}, \bibinfo
  {author} {\bibfnamefont {X.}~\bibnamefont {Zhang}}, \bibinfo {author}
  {\bibfnamefont {Y.}~\bibnamefont {Su}}, \bibinfo {author} {\bibfnamefont
  {X.}~\bibnamefont {Wang}}, \bibinfo {author} {\bibfnamefont {Z.}~\bibnamefont
  {Liu}}, \bibinfo {author} {\bibfnamefont {Y.}~\bibnamefont {Wang}}, \bibinfo
  {author} {\bibfnamefont {R.}~\bibnamefont {Zhu}}, \bibinfo {author}
  {\bibfnamefont {Y.}~\bibnamefont {Wang}}, \bibinfo {author} {\bibfnamefont
  {W.}~\bibnamefont {Liu}}, \ and\ \bibinfo {author} {\bibfnamefont
  {J.}~\bibnamefont {Tang}},\ }\href@noop {} {\bibfield  {journal} {\bibinfo
  {journal} {Applied Physics Letters}\ }\textbf {\bibinfo {volume} {98}},\
  \bibinfo {pages} {192510} (\bibinfo {year} {2011}{\natexlab{b}})}\BibitemShut
  {NoStop}%
\bibitem [{\citenamefont {Mandal}\ \emph {et~al.}(2010)\citenamefont {Mandal},
  \citenamefont {Sundaresan}, \citenamefont {Rao}, \citenamefont {Iyo},
  \citenamefont {Shirage}, \citenamefont {Tanaka}, \citenamefont {Simon},
  \citenamefont {Pralong}, \citenamefont {Lebedev}, \citenamefont {Caignaert},\
  and\ \citenamefont {Raveau}}]{Mandal2010}%
  \BibitemOpen
  \bibfield  {author} {\bibinfo {author} {\bibfnamefont {P.}~\bibnamefont
  {Mandal}}, \bibinfo {author} {\bibfnamefont {A.}~\bibnamefont {Sundaresan}},
  \bibinfo {author} {\bibfnamefont {C.~N.~R.}\ \bibnamefont {Rao}}, \bibinfo
  {author} {\bibfnamefont {A.}~\bibnamefont {Iyo}}, \bibinfo {author}
  {\bibfnamefont {P.~M.}\ \bibnamefont {Shirage}}, \bibinfo {author}
  {\bibfnamefont {Y.}~\bibnamefont {Tanaka}}, \bibinfo {author} {\bibfnamefont
  {C.}~\bibnamefont {Simon}}, \bibinfo {author} {\bibfnamefont
  {V.}~\bibnamefont {Pralong}}, \bibinfo {author} {\bibfnamefont {O.~I.}\
  \bibnamefont {Lebedev}}, \bibinfo {author} {\bibfnamefont {V.}~\bibnamefont
  {Caignaert}}, \ and\ \bibinfo {author} {\bibfnamefont {B.}~\bibnamefont
  {Raveau}},\ }\href@noop {} {\bibfield  {journal} {\bibinfo  {journal}
  {Physical Review B}\ }\textbf {\bibinfo {volume} {82}},\ \bibinfo {pages}
  {100416} (\bibinfo {year} {2010})}\BibitemShut {NoStop}%
\bibitem [{\citenamefont {a.K. Azad}\ \emph {et~al.}(2005)\citenamefont {a.K.
  Azad}, \citenamefont {Mellerg{\aa}rd}, \citenamefont {Eriksson},
  \citenamefont {Ivanov}, \citenamefont {Yunus}, \citenamefont {Lindberg},
  \citenamefont {Svensson},\ and\ \citenamefont {Mathieu}}]{Azad2005}%
  \BibitemOpen
  \bibfield  {author} {\bibinfo {author} {\bibnamefont {a.K. Azad}}, \bibinfo
  {author} {\bibfnamefont {a.}~\bibnamefont {Mellerg{\aa}rd}}, \bibinfo
  {author} {\bibfnamefont {S.-G.}\ \bibnamefont {Eriksson}}, \bibinfo {author}
  {\bibfnamefont {S.}~\bibnamefont {Ivanov}}, \bibinfo {author} {\bibfnamefont
  {S.}~\bibnamefont {Yunus}}, \bibinfo {author} {\bibfnamefont
  {F.}~\bibnamefont {Lindberg}}, \bibinfo {author} {\bibfnamefont
  {G.}~\bibnamefont {Svensson}}, \ and\ \bibinfo {author} {\bibfnamefont
  {R.}~\bibnamefont {Mathieu}},\ }\href@noop {} {\bibfield  {journal} {\bibinfo
   {journal} {Materials Research Bulletin}\ }\textbf {\bibinfo {volume} {40}},\
  \bibinfo {pages} {1633} (\bibinfo {year} {2005})}\BibitemShut {NoStop}%
\bibitem [{\citenamefont {Pomiro}\ \emph {et~al.}(2016)\citenamefont {Pomiro},
  \citenamefont {S\'anchez}, \citenamefont {Cuello}, \citenamefont {Maignan},
  \citenamefont {Martin},\ and\ \citenamefont {Carbonio}}]{Pomiro2016}%
  \BibitemOpen
  \bibfield  {author} {\bibinfo {author} {\bibfnamefont {F.}~\bibnamefont
  {Pomiro}}, \bibinfo {author} {\bibfnamefont {R.~D.}\ \bibnamefont
  {S\'anchez}}, \bibinfo {author} {\bibfnamefont {G.}~\bibnamefont {Cuello}},
  \bibinfo {author} {\bibfnamefont {A.}~\bibnamefont {Maignan}}, \bibinfo
  {author} {\bibfnamefont {C.}~\bibnamefont {Martin}}, \ and\ \bibinfo {author}
  {\bibfnamefont {R.~E.}\ \bibnamefont {Carbonio}},\ }\href@noop {} {\bibfield
  {journal} {\bibinfo  {journal} {submitted}\ } (\bibinfo {year}
  {2016})}\BibitemShut {NoStop}%
\bibitem [{\citenamefont {Murtazaev}\ \emph {et~al.}(2005)\citenamefont
  {Murtazaev}, \citenamefont {Kamilov},\ and\ \citenamefont
  {Ibaev}}]{Murtazaev05LTP}%
  \BibitemOpen
  \bibfield  {author} {\bibinfo {author} {\bibfnamefont {A.~K.}\ \bibnamefont
  {Murtazaev}}, \bibinfo {author} {\bibfnamefont {I.~K.}\ \bibnamefont
  {Kamilov}}, \ and\ \bibinfo {author} {\bibfnamefont {Z.~G.}\ \bibnamefont
  {Ibaev}},\ }\href@noop {} {\bibfield  {journal} {\bibinfo  {journal} {Low
  Temperature Physics}\ }\textbf {\bibinfo {volume} {31}},\ \bibinfo {pages}
  {139} (\bibinfo {year} {2005})}\BibitemShut {NoStop}%
\bibitem [{\citenamefont {Restrepo-Parra}\ \emph {et~al.}(2010)\citenamefont
  {Restrepo-Parra}, \citenamefont {Bedoya-Hincapié}, \citenamefont {Jurado},
  \citenamefont {Riano-Rojas},\ and\ \citenamefont
  {Restrepo}}]{RestrepoParra10JMMM}%
  \BibitemOpen
  \bibfield  {author} {\bibinfo {author} {\bibfnamefont {E.}~\bibnamefont
  {Restrepo-Parra}}, \bibinfo {author} {\bibfnamefont {C.}~\bibnamefont
  {Bedoya-Hincapié}}, \bibinfo {author} {\bibfnamefont {F.}~\bibnamefont
  {Jurado}}, \bibinfo {author} {\bibfnamefont {J.}~\bibnamefont {Riano-Rojas}},
  \ and\ \bibinfo {author} {\bibfnamefont {J.}~\bibnamefont {Restrepo}},\
  }\href@noop {} {\bibfield  {journal} {\bibinfo  {journal} {Journal of
  Magnetism and Magnetic Materials}\ }\textbf {\bibinfo {volume} {322}},\
  \bibinfo {pages} {3514 } (\bibinfo {year} {2010})}\BibitemShut {NoStop}%
\bibitem [{\citenamefont {Restrepo-Parra}\ \emph {et~al.}(2011)\citenamefont
  {Restrepo-Parra}, \citenamefont {Salazar-Enríquez}, \citenamefont
  {Londoño-Navarro}, \citenamefont {Jurado},\ and\ \citenamefont
  {Restrepo}}]{RestrepoParra11JMMM}%
  \BibitemOpen
  \bibfield  {author} {\bibinfo {author} {\bibfnamefont {E.}~\bibnamefont
  {Restrepo-Parra}}, \bibinfo {author} {\bibfnamefont {C.}~\bibnamefont
  {Salazar-Enríquez}}, \bibinfo {author} {\bibfnamefont {J.}~\bibnamefont
  {Londoño-Navarro}}, \bibinfo {author} {\bibfnamefont {J.}~\bibnamefont
  {Jurado}}, \ and\ \bibinfo {author} {\bibfnamefont {J.}~\bibnamefont
  {Restrepo}},\ }\href@noop {} {\bibfield  {journal} {\bibinfo  {journal}
  {Journal of Magnetism and Magnetic Materials}\ }\textbf {\bibinfo {volume}
  {323}},\ \bibinfo {pages} {1477 } (\bibinfo {year} {2011})}\BibitemShut
  {NoStop}%
\bibitem [{\citenamefont {Hashimoto}(1963)}]{Hashimoto63JPSJ}%
  \BibitemOpen
  \bibfield  {author} {\bibinfo {author} {\bibfnamefont {T.}~\bibnamefont
  {Hashimoto}},\ }\href@noop {} {\bibfield  {journal} {\bibinfo  {journal}
  {Journal of the Physical Society of Japan}\ }\textbf {\bibinfo {volume}
  {18}},\ \bibinfo {pages} {1140} (\bibinfo {year} {1963})}\BibitemShut
  {NoStop}%
\bibitem [{\citenamefont {{Moriya}}(1963)}]{Bertaut}%
  \BibitemOpen
  \bibfield  {author} {\bibinfo {author} {\bibfnamefont {T.}~\bibnamefont
  {{Moriya}}},\ }\href@noop {} {\emph {\bibinfo {title} {Magnetism III}}},\
  \bibinfo {edition} {editing by g. t. rado and h. suhl}\ ed.\ (\bibinfo
  {publisher} {Academic Press},\ \bibinfo {address} {New York},\ \bibinfo
  {year} {1963})\BibitemShut {NoStop}%
\bibitem [{\citenamefont {C.}\ \emph {et~al.}(1959)\citenamefont {C.},
  \citenamefont {P.},\ and\ \citenamefont {J.}}]{Sherwood1959}%
  \BibitemOpen
  \bibfield  {author} {\bibinfo {author} {\bibfnamefont {S.~R.}\ \bibnamefont
  {C.}}, \bibinfo {author} {\bibfnamefont {R.~J.}\ \bibnamefont {P.}}, \ and\
  \bibinfo {author} {\bibfnamefont {W.~H.}\ \bibnamefont {J.}},\ }\href@noop {}
  {\bibfield  {journal} {\bibinfo  {journal} {J. Appl. Phys.}\ }\textbf
  {\bibinfo {volume} {30}},\ \bibinfo {pages} {217} (\bibinfo {year}
  {1959})}\BibitemShut {NoStop}%
\bibitem [{\citenamefont {Hornreich}\ \emph {et~al.}(1976)\citenamefont
  {Hornreich}, \citenamefont {Shtrikman}, \citenamefont {Wanklyn},\ and\
  \citenamefont {Yaeger}}]{Hornreich76PRB}%
  \BibitemOpen
  \bibfield  {author} {\bibinfo {author} {\bibfnamefont {R.~M.}\ \bibnamefont
  {Hornreich}}, \bibinfo {author} {\bibfnamefont {S.}~\bibnamefont
  {Shtrikman}}, \bibinfo {author} {\bibfnamefont {B.~M.}\ \bibnamefont
  {Wanklyn}}, \ and\ \bibinfo {author} {\bibfnamefont {I.}~\bibnamefont
  {Yaeger}},\ }\href@noop {} {\bibfield  {journal} {\bibinfo  {journal} {Phys.
  Rev. B}\ }\textbf {\bibinfo {volume} {13}},\ \bibinfo {pages} {4046}
  (\bibinfo {year} {1976})}\BibitemShut {NoStop}%
\bibitem [{\citenamefont {Treves}(1965)}]{Treves65JAP}%
  \BibitemOpen
  \bibfield  {author} {\bibinfo {author} {\bibfnamefont {D.}~\bibnamefont
  {Treves}},\ }\href@noop {} {\bibfield  {journal} {\bibinfo  {journal}
  {Journal of Applied Physics}\ }\textbf {\bibinfo {volume} {36}},\ \bibinfo
  {pages} {1033} (\bibinfo {year} {1965})}\BibitemShut {NoStop}%
\bibitem [{\citenamefont {Yuan}\ \emph {et~al.}(2012)\citenamefont {Yuan},
  \citenamefont {Tang}, \citenamefont {Sun},\ and\ \citenamefont
  {Xu}}]{Yuan2012}%
  \BibitemOpen
  \bibfield  {author} {\bibinfo {author} {\bibfnamefont {X.~P.}\ \bibnamefont
  {Yuan}}, \bibinfo {author} {\bibfnamefont {Y.}~\bibnamefont {Tang}}, \bibinfo
  {author} {\bibfnamefont {Y.}~\bibnamefont {Sun}}, \ and\ \bibinfo {author}
  {\bibfnamefont {M.~X.}\ \bibnamefont {Xu}},\ }\href@noop {} {\bibfield
  {journal} {\bibinfo  {journal} {J. Appl. Phys.}\ } (\bibinfo {year}
  {2012})}\BibitemShut {NoStop}%
\bibitem [{\citenamefont {Sahu}\ \emph {et~al.}(2007)\citenamefont {Sahu},
  \citenamefont {Serrao}, \citenamefont {RAy}, \citenamefont {Waghmare},\ and\
  \citenamefont {Rao}}]{Sahu2007}%
  \BibitemOpen
  \bibfield  {author} {\bibinfo {author} {\bibfnamefont {J.~R.}\ \bibnamefont
  {Sahu}}, \bibinfo {author} {\bibfnamefont {C.~R.}\ \bibnamefont {Serrao}},
  \bibinfo {author} {\bibfnamefont {N.}~\bibnamefont {RAy}}, \bibinfo {author}
  {\bibfnamefont {U.~V.}\ \bibnamefont {Waghmare}}, \ and\ \bibinfo {author}
  {\bibfnamefont {C.~N.}\ \bibnamefont {Rao}},\ }\href@noop {} {\bibfield
  {journal} {\bibinfo  {journal} {J. Mater. Chem.}\ } (\bibinfo {year}
  {2007})}\BibitemShut {NoStop}%
\end{thebibliography}

%
\end{document}